\documentclass[fleqn,usenatbib,useAMS]{mnras}

\usepackage[T1]{fontenc}
\usepackage{ae,aecompl}
\usepackage{graphicx}	% Including figure files
\usepackage{amsmath}	% Advanced maths commands
\usepackage{amssymb}	% Extra maths symbols
\usepackage[british]{babel}
\usepackage{multicol}        % Multi-column entries in tables
\usepackage{bm}		% Bold maths symbols, including upright Greek
\usepackage{pdflscape}	% Landscape pages

\title[SFDM in clusters of galaxies]{Scalar field dark matter in
clusters of galaxies}

\author[T. Bernal, V.H. Robles and T. Matos]
       {Tula Bernal,$^{1}$\thanks{E-mail: tbernal@fis.cinvestav.mx}
       \thanks{Present address:
       Departamento de F\'{\i}sica, Instituto Nacional de
       Investigaciones Nucleares, AP 18-1027, Ciudad de M\'{e}xico
       11801, M\'{e}xico}
       Victor H. Robles$^{2}$\thanks{E-mail: vrobles@uci.edu}
       and Tonatiuh Matos$^{1}$\thanks{E-mail: tmatos@fis.cinvestav.mx}
       \thanks{Part of the Instituto Avanzado de Cosmolog\'ia
	   (IAC) collaboration (http://www.iac.edu.mx/)}
       \\
       $^{1}$Departamento de F\'{\i}sica, Centro de Investigaci\'on y de
       Estudios Avanzados del Instituto Polit\'ecnico Nacional, AP
       14-740,\\Ciudad de M\'exico 07000, M\'exico\\
       $^{2}$Department of Physics and Astronomy, University of
       California, Irvine, 4129 Frederick Reines Hall, Irvine, CA 92697,
       USA
       }

\date{Accepted 2017 March 14. Received 2017 February 16; in original form 2016 October 12}
\pubyear{2017}

\begin{document}
\label{firstpage}
\pagerange{\pageref{firstpage}--\pageref{lastpage}}
\maketitle

%%%%% ABSTRACT %%%%%%%
\begin{abstract}
  One alternative to the cold dark matter (CDM) paradigm is the scalar
field dark matter (SFDM) model, which assumes dark matter is a spin-0
ultra-light scalar field (SF) with a typical mass $m\sim10^{-22}\mathrm{eV}/
c^2$ and positive self-interactions. Due to the ultra-light boson mass,
the SFDM could form Bose-Einstein condensates (BEC) in the very early
Universe, which are interpreted as the dark matter haloes. Although
cosmologically the model behaves as CDM, they differ at small scales:
SFDM naturally predicts fewer satellite haloes, cores in dwarf galaxies
and the formation of massive galaxies at high redshifts. The ground
state (or BEC) solution at zero temperature suffices to describe
low-mass galaxies but fails for larger systems. A possible solution is
adding finite-temperature corrections to the SF potential which allows
combinations of excited states. In this work, we test the
finite-temperature multistate SFDM solution at galaxy cluster scales and
compare our results with the Navarro-Frenk-White (NFW) and BEC profiles. We achieve this by
fitting the mass distribution of 13 \textit{Chandra} X-ray clusters of
galaxies, excluding the region of the brightest cluster galaxy. We show
that the SFDM model accurately describes the clusters' DM mass
distributions offering an equivalent or better agreement than the NFW
profile. The complete disagreement of the BEC model with the data is
also shown. We conclude that the theoretically motivated multistate SFDM
profile is an interesting alternative to empirical profiles and
ad hoc fitting-functions that attempt to couple the asymptotic
NFW decline with the inner core in SFDM.
\end{abstract}


\section{Introduction}
\label{introduction}

  One of the biggest challenges of cosmology and astrophysics is to
understand how galaxies and clusters of galaxies were formed and
evolved. In the context of General Relativity, it is known that without
the assumption of a cold dark matter (CDM) component it is difficult to
explain the observed anisotropies in the cosmic microwave background (CMB)
radiation, the large-scale structure formation in the Universe, the
galactic formation processes and gravitational lenses of distant
objects, among others.  Moreover, adding a positive cosmological
constant $\Lambda$ can account for the accelerated expansion of the
Universe. These components, along with the baryonic matter, form the
current paradigm explaining the dynamics of the Universe, known as the
$\Lambda$CDM or standard model. Current
observations from the {\it{Planck}} mission set the contribution of the
baryonic matter to the total matter-energy density of the Universe to
$\sim 5\%$, meanwhile the CDM is $\sim 26\%$ and $\Lambda$ or dark
energy is $\sim 69\%$ \citep{Planck:2016}.

  From CDM $N$-body simulations of structure formation, we know that CDM
clusters form haloes with the universal Navarro-Frenk-White (NFW)
density profile \citep{nfw-1997}, which is proportional to $r^{-1}$ (a
`cuspy' profile) for small radii $r$ and to $r^{-3}$ for large radii
(cf. equation~\ref{rho-nfw}).  Even though the CDM paradigm is very
successful at reproducing the large-scale observations, recent DM-only
simulations are consistent with a `cuspy' profile \citep{navarro10},
meanwhile high-resolution observations of dark matter dominated systems,
such as low surface brightness (LSB) galaxies \citep{deblok:2001}
and dwarf spheroidal (dSph) galaxies \citep{oh:2011, walker:2011,
penarrubia:2012}, suggest a constant central density or `core' profile
($\rho \sim r^{-0.2}$). This discrepancy is known as the `cusp-core
problem'.

  In addition to the cusp-core issue, the CDM paradigm faces other
challenges on small scales: it predicts more massive satellite galaxies
around Milky Way like galaxies that have not been observed
\citep{boylan:2011,sawala:2012}; it fails to reproduce the phase-space
distribution of satellites around the Milky Way and Andromeda galaxies
\citep{pawlowski:2012,ibata:2013,ibata:2014} and the internal dynamics
in tidal dwarf galaxies \citep{gentile:2007,kroupa:2012}. Another
potential difficulty may lie in the early formation of large galaxies:
large systems are formed hierarchically through mergers of small
galaxies that collapsed earlier, but recent observations have found
various massive galaxies at very high redshifts \citep{caputi:2015}; it
remains to be seen whether such rapid formation could represent a
problem for the CDM model.

  An active field of research to solve the above issues takes into
account the baryonic physics \citep{Navarro:1996,governato:2010,
deblok:2010,maccio:2012,stinson:2012,DiCintio:2014,Pontzen:2014,
Onorbe:2015,Chan:2015}, so the NFW universal profile is not expected to
hold exactly once the simulations include the baryonic matter. Those
have shown that a core profile can be obtained in simulations of dwarf
galaxies in CDM if they include a bursty and continuous star formation
rate \citep{pontzen:2012,governato:2012,Teyssier:2013,Madau:2014,
DiCintio:2014,Onorbe:2015}.  However, it is difficult to explain flat
density profiles with these mechanisms in galaxies with masses smaller
than $\sim 10^{6.5} M_\odot$ \citep{penarrubia:2012,
Garrison-Kimmel:2013,Pontzen:2014,Chan:2015} and, possibly, in some
LSB galaxies \citep[see e.g.][]{kuzio:2011}. Very recent simulations
have shown that baryons could explain the small abundance of low-mass
galaxies \citep{FIRE:2017}; it remains therefore uncertain whether the
baryonic processes (star formation, supernova explosions, active
galactic nuclei, stellar winds, etc.) and their effect on the dark
matter haloes are enough to account for the discrepancies.

  There are some empirical density profiles proposed in order to describe
the density distribution after accounting for the baryonic component,
for instance, the Burkert profile \citep{burkert:1995} or the generalized
NFW profile \citep{Zhao:1996}. However, these parametrizations, albeit
useful, lack of theoretical support; it is therefore interesting to
explore alternative dark matter models from which we can derive a density
profile that agrees with a broad range of observations.

  A different approach is to find models trying to replace the dark
matter hypothesis with a modified gravity law to explain the
observations at different scales \citep[see e.g.][]{defelice:2010,
mendoza11, bernal11, Famaey:2012, joyce:2015}.  There are some gravity
models capable of reproducing the observations of galaxy clusters and
consistent at the galactic level also \citep{khoury:2015,Bernal:2015};
more work is being developed in this direction.
   
  Some of the DM alternatives to the CDM paradigm are warm dark matter,
self-interacting dark matter \citep{spergel:2000,yoshida:2000,dave:2001,
zavala:2009,navarro10,kuzio:2011,Elbert:2015,Robles:2017}, and scalar
field dark matter \citep[SFDM,][]{Sin:1992,Ji:1994,Lee:1995}. In this work we will
focus our study on the SFDM alternative in the mass range of galaxy
clusters.

  The scalar fields (SF) as dark matter were first elucidated by
\citet{Ruffini:1983}; since then the idea was rediscovered using
different names \citep[see e.g.][]{Membrado:1989,Spergel:1989,Sin:1992,
Ji:1994,Lee:1995,guzman-matos:2000,Sahni:1999,Peebles:2000,Goodman:2000,
Matos-Urena:2000,matos-urena01,hu:2000,Wetterich:2001,Arbey:2001a,
harko07,Vazquez-Magana:2008,Woo:2009,Lundgren:2010,Bray:2010,
Marsh-Ferreira:2010,Robles:2013,Schive:2014,Calabrese:2016}, among
others, and more recently by \cite*{Ostriker:2016}. However, the first
systematic study of the cosmological behaviour started by
\citet{Guzman:1999,guzman-matos:2000,Matos:1999,Matos-Urena:2000}. Other
systematic studies were performed by \citet{Arbey:2001a,Arbey:2001b} and
more recently by \citet{Marsh-Ferreira:2010,Schive:2014,Marsh:2015b}.

  In the SFDM model, the DM is a SF of spin-0 interaction, 
motivated by the well-known fundamental interactions of spin-1 or -2,
being the spin-0 the simplest one. The model considers a spin-0 SF of
a very small mass (typically $\sim 10^{-22} \mathrm{eV}/
c^2$) as the dark matter, in such a way that it is possible to form
Bose-Einstein Condensates (BEC) at cosmological scales, hence behaving as
CDM at large scales. The model has been widely explored by
many authors, named simply SFDM
\citep{Matos-Urena:2000,matos-urena01,alcubierre:2002a,alcubierre:2002b,
matos-urena07,Matos:2009,suarez:2011,Magana:2012,robles:2012,Robles:2013,
Robles-lensing:2013,Martinez-Medina:2014hca,Suarez:2014,Martinez:2015,
Robles:2015,Matos-Robles:2016}. As the SFDM model gained interest,
several authors gave it different names in the literature; however, we
emphasize that the core idea described above remains unchanged. Some of
the names are: fuzzy \citep{hu:2000}, wave \citep{bray:2012,Schive:2014,
Schive:2014hza}, BEC \citep{harko07} or
ultra-light axion \citep{Marsh-Ferreira:2010,hlozek:2015} dark matter.
Additionally, most of these works assume the SFDM is at zero temperature,
implying the SFDM ultra-light bosons occupy the ground state only. Some
authors have explored thoroughly the possibility that the axion from quantum
chromodynamics \citep[see e.g.][]{Peccei-Quinn:1977,Frieman:1995,Fox:2004} and string
theory \citep[see e.g.][]{Arvanitaki:2010} is the DM of the Universe
\citep[see][for a review]{Marsh:2015b}. The axion is a spin-0 particle
with small mass ($\sim 10^{-5} \mathrm{eV}/c^2$) and weak
self-interaction, and it has been proposed that axion DM can form BECs
during the radiation-dominated era \citep{Sikivie:2009,Erken:2012}.
However, it has not been proved that such axion BECs can be the DM haloes
of the galaxies and clusters of galaxies \citep[see e.g.][]
{Chavanis:2016}.

  The mentioned works on SFDM have shown that the model can account for
the CDM discrepancies for a typical mass of the SF of $m\sim10^{-22}
\mathrm{eV}/c^2$.  Moreover, it was found that for small galaxies, the
ground state solution is sufficient to reproduce current observations
\citep{harko07,robles:2012}. However, as the galaxies become larger,
temperature corrections to the SFDM potential are needed in order to
obtain a solution that can account for the contribution of excited
states and agree with the observational data \citep{Robles:2013}. From
finite-temperature quantum field theory it is possible to obtain
one-loop temperature corrections for the SF potential
\citep{kolb:1994}. Following this approach, \cite{Robles:2013} found an
approximate analytic solution to the field equations and derived a 
density profile that allows combinations of excited states of the SF.
That solutions correspond to self-gravitating systems of SFDM, which are
interpreted as dark matter haloes; they are also called multistate
haloes. Such finite-temperature analytic solution has successfully
described galactic systems \citep{Robles:2013,Bernal:2017}. In this
article, we explore the galaxy clusters regime; our goal is to assess the
viability of the SFDM model in these systems. In particular, we test the
finite-temperature solution and compare it with the results of applying
the NFW density profile to the same observations. We do so by fitting
the SFDM mass profile to the mass distribution of 13 galaxy clusters
from \citet{Vikhlinin:2006} and obtain that the multistate solution, and
hence the finite-temperature SFDM model, is successful at reproducing
the clusters mass regime, offering an alternative profile to describe
the massive systems but with theoretical support, compared to the usual
approach of fitting empirical profiles to observational data.
Additionally, we demonstrate that the solution of a self-gravitating
configuration at zero temperature and in the Thomas-Fermi limit, where
the SF self-interactions dominate the SFDM potential \citep{harko07}, is
incapable of fitting the galaxy clusters at all radii, which strongly
favours the usage of excited states. 

  The paper is organized as follows: In Section~\ref{sfdm-model}, we
briefly review the SFDM model, mentioning the BEC dark matter at zero
temperature in the TF limit and the analytic multistate SFDM
model. In Section~\ref{cluster-profiles}, we provide the analysis tools
required to derive the cluster mass profiles from the X-ray
observations and perform the fits to the DM component once baryons are
subtracted. In Section~\ref{discussion}, we show and discuss the results
of the comparison between the three DM profiles: the SFDM ground state at
zero temperature (BEC), the multistate SFDM halo with
temperature-corrections and the NFW profile, representative of the CDM
model. Finally, in Section~\ref{summary} we summarize our results and
include our conclusions.

\section{SFDM model}
\label{sfdm-model}

  In the SFDM model, the DM is considered to be a real SF with
positive self-interactions that forms BEC
`drops' \citep{Sin:1992,Ji:1994,Lee:1995,guzman-matos:2000}. Recently,
this model has received more attention due to the natural solution to
the standard CDM problems and the agreement with observations
\citep{Suarez:2014}. The cosmological behaviour of the SFDM was first
investigated by \citet{matos-urena01}, who showed that the BEC
configurations were formed at critical condensation temperatures of TeV,
implying that these BEC-DM haloes could have been formed very early in
the Universe. Furthermore, \citet{matos-urena01} found that the SF power
spectrum has an intrinsic cut-off that prevents the growth of small
haloes: a dark matter boson with mass $m \sim 10^{-22} \mathrm{eV}/c^2$
reduces the abundance of haloes with masses $M \sim 10^{8}M_{\odot}$
observed today, which substantially reduces the amount of satellite
haloes around a Milky Way like galaxy. This last result represents an
advantage with respect to the CDM model, naturally predicting few
substructure around big galaxies \citep[see also][]{Urena:2015}, and has
been confirmed using numerical simulations \citep{Schive:2014} and
semi-analytic models \citep{hlozek:2015}. Studies of the effect of tidal
forces on substructure embedded in a SFDM potential find that the same
mass can explain the long-lived stellar clump in Ursa Minor
\citep{Lora:2012} and the survivability of small satellite SFDM haloes
orbiting around a Milky Way like host galaxy \citep{Robles:2015}.

 It was also first shown by \citet{matos-urena01} that the SF behaves as
dust at cosmological scales and that the CMB as well as the
mass power spectrum (MPS) are similar to those
found in the CDM paradigm at large scales, thus the SFDM model
reproduces the cosmological observations as well as CDM, up to
linear-order perturbations \citep[see][for better resolution plots of
the CMB and MPS]{rguez-montoya:2010, hlozek:2015,schive:2015}
\citep[see also][]{Matos:2009,harko:2011,suarez:2011, chavanis:2011,
Magana:2012,Schive:2014,suarez:2015}.

  Another important result is that the model naturally avoids cuspy
haloes, a consequence of the wave properties of the SF and the
Heisenberg uncertainty principle acting on kpc scales; both properties
prevent the DM density from growing indefinitely in a small region
\citep{hu:2000}, producing a constant density in the centre of the DM
distribution. Interestingly, a mass of $m \sim 10^{-22} \mathrm{eV}/c^2$
results in a $\sim$kpc core size similar to what is suggested by
observations of dSph galaxies \citep{Chen-Schive:2016,
Gonzalez-Marsh:2016}.  

  Using numerical simulations, \citet{alcubierre:2002a,alcubierre:2002b}
found that the critical mass of collapse for the SF is given
by $M_\mathrm{crit}$ $\sim$ $0.1 m^2_\mathrm{Pl}/m$, where $m_\mathrm{Pl}$
is the Planck mass.  For the ultra-light boson mass, $m$ $\sim$ $10^{-22}
\mathrm{eV}/c^2$, the critical mass of stability is $M_\mathrm{crit}$
$\sim$ $10^{12}M_\odot$. Thus, some massive haloes could be born close to
the critical mass and induce rapid formation of large galaxies at high
redshifts. Therefore, another main prediction of the SFDM model is the
existence of massive galaxies at high redshifts due to the early halo
formation \citep[see also][]{Magana:2012}. In order to give a
quantitative estimation we need cosmological simulations that can address
the non-linear regime of SFDM halo formation. 

  Finally, from stability studies of SFDM configurations,
\cite{Colpi:1986, Gleiser:1988,Sin:1992,balakrishna:1998,
guzman-matos:2000} have found that very massive systems might be
unstable if they have masses above the critical one, suggesting that
structures like clusters of galaxies, with masses of $M \approx 10^{13}
- 10^{15} M_\odot$ are formed by mergers, just like in the standard CDM
paradigm.  Nevertheless, for SFDM configurations that include excited
states \citep{Seidel-Suen:1990,Hawley:2003,Urena:2009,BernalA:2010,
urena-bernal10}, the resulting configuration can have a larger mass and
then migrate to lower energy states until it reaches a new equilibrium 
configuration through mass loss (usually called gravitational cooling). 

  \cite{urena-bernal10} showed it is possible to have stable 
configurations with multistates that do not decay completely to the
ground state provided the fraction of the total mass in each of the
different states is similar, in particular they found the threshold for
stability for a multistate halo formed by the ground+first excited
states. They took the ratio $\eta$ of the mass in the excited state,
$M_j$, with respect to the total mass in the ground state, $M_1$, and
found that values of $\eta= M_j/M_1 \lesssim 1.3$ yield stable multistate
haloes.  Given that clusters are formed by mergers of different haloes,
the ultra-light bosons are expected to mix and occupy different states as
they distribute along the spatial extent of the cluster; the final state
in general will depend on the merger history of each cluster.  We pursue
the multistate halo merger idea and show that a good approximation to the
total mass profile can be obtained if we account for the superposition of
individual states. The final mass profile presents small oscillations at
large radii as a result of adding the excited states; interestingly,
similar oscillations have been noted in numerical simulations that
explore the formation of smaller SFDM haloes by merging haloes with
different mass ratios \citep{Schive:2014hza,Guzman:2016,Schwabe:2016}. We
believe the origin of such oscillations is the interference of the
different intrinsic modes that compose the total wavefunction, and using
the profile from the SFDM with temperature corrections, which includes
the combination of states, we can in fact describe such oscillations at
large radii. In this way it is not necessary to invoke a two-component
empirical profile tuned to a NFW profile in the outskirts and a core
profile for small radii \citep{Schive:2014,Schive:2014hza,
Marsh-Pop:2015}.  To confirm our claim it is necessary to develop a mode
decomposition of the total halo mass profiles in a numerical simulation,
but we leave a detailed comparison study for a future work.

\subsection{BEC haloes at \textit{T}=0 in the Thomas-Fermi limit}

  There are numerous works that study bosons at zero temperature. As its
Compton wavelength is large, the self-gravitating SFDM haloes have large
occupation numbers; at $T$=0 practically all the bosons are in the
ground state. Applying the \citet{Bogolyubov:1947} approximation, the
BEC configurations can be described by a mean-field classical SF
representing the ground state, neglecting the excited states
contribution due to the zero-temperature hypothesis.  In the Newtonian
limit, the BEC is described by the Gross-Pitaevskii (GP) equation (that
describes the ground state of a BEC at $T$=0) and the Poisson equation.
\citet{harko07} applied the Thomas-Fermi (TF) limit, in which the
self-interactions of the SF $\psi$ dominate in the SFDM potential, such
that the mass term ($\sim \psi^2$) can be neglected and the resulting
potential goes as $V(\psi)\sim\lambda \psi^4$.  They used this limit to
derive the solution to the GP-Poisson system, which reads:
\begin{equation}
	\nabla^2 \rho_\mathrm{BEC}(r) + k^2 \rho_\mathrm{BEC}(r) = 0 .
\label{eq-1}
\end{equation}
The solution found is
\begin{equation}
	\rho_\mathrm{BEC}(r) = \rho^0_\mathrm{BEC} \frac{\sin(kr)}{kr} ,
\label{zero-dens}
\end{equation}
where $\rho^0_\mathrm{BEC} := \rho_\mathrm{BEC}(0)$ is the central
density and $k$ is a parameter. To exclude non-physical negative
densities, the solution has a cut-off to a bound halo radius $\hat{R}$
determined by $\rho_\mathrm{BEC}(\hat R)=0$, in this case for $k := \pi/ 
\hat R$. This expression fixes the radius of the BEC halo to
\citep{harko07}
\begin{equation}
 \hat R= \pi \sqrt{\frac{\hbar^{2} a}{G m^{3}}},
\label{R-hat}
\end{equation}
where $\hbar$ is the reduced Planck's constant, $G$ the gravitational 
constant, $m$ the mass of the SF particle and $a$ its scattering 
length. Such equation is related to the coupling constant $\lambda$ by
$\lambda=4 \pi \hbar^2 a/m$. Using this solution, it was shown that it
is possible to fit the flat rotation curves of small galaxies but only
up to the cut-off radius $\hat{R}$; for $r>\hat{R}$, the rotation curves 
follow the standard Keplerian law, and the velocity profile decreases
very quickly after its maximum value \citep{harko07,robles:2012,
Guzman:2015}; also the BEC haloes in the TF limit have shown to be
unstable \citep{Guzman:2013}. Thus, the approximation for the fully
condensed system at $T$=0 works well to fit only galaxy haloes of sizes
around $5-7$ kpc, but struggles to model larger galaxies, which in
contrast can be easily explained with excited states as their effect is
to flatten the rotation curves at outer radii
\citep{urena-bernal10,Robles:2013,Martinez-Medina:2014hca,Bernal:2017}.

  Another difficulty of the profile at zero temperature is that different
values for the parameter $\hat{R}$ are needed for galaxies of different
sizes (cf. equation~\ref{R-hat}), which means that the intrinsic properties
of the DM particle ($a$, $m$) vary from galaxy to galaxy. Moreover,
we show in Section~\ref{discussion} that this model does not fit the
dynamical masses of the galaxy clusters analyzed in the present work.

\subsection{Finite-temperature multistate SFDM haloes}
\label{finite-temperature-models}

 A finite-temperature SF scenario that has been shown to solve the
discrepancies in the rotation curves of galaxies was proposed by
\citet{Robles:2013}. They considered a self-interacting spin-0 real SF
embedded in a thermal bath of temperature $T$; the first-order
contribution of the temperature to the SF potential is given by the 
temperature corrections up to one-loop in perturbations
\citep{kolb:1994}:
\begin{equation}
	V(\psi) = - \frac{1}{2} m^2 \psi^2 + \frac{1}{4} \lambda \psi^4 +
	\frac{1}{8} \lambda \psi^2 T^2 - \frac{\pi^2}{90} T^4 \, ,
\label{eq:ft-potential}
\end{equation}
in units $\hbar=1$ for the reduced Planck's constant, $c=1$ for the
speed of light and $k_\mathrm{B}=1$ for the Boltzmann constant. Such SF
potential has a $Z_2$ symmetry for high temperatures in the very early
Universe. As the temperature decreases with the expansion of the
Universe, the SF undergoes a spontaneous symmetry breaking and the system
rolls down to a new minimum of the potential, where the SF perturbations
can grow and form the early DM haloes with different equilibrium
temperatures depending on the formation time of each halo. Assuming the
field is in the minimum of the potential, the authors derived an exact
analytic spherically symmetric static solution for SF haloes in the
Newtonian limit. As the mass density is proportional to $\psi^2$, the
finite-temperature SFDM density is given by \citep[see][for more details
on the calculations]{Robles:2013}
\begin{equation}
	\rho^j_\mathrm{SFDM}(r) = \rho^j_0 \left[ \frac{\sin(k_j r)}
	{(k_j r)} \right]^2, 
\label{T-density}
\end{equation}
where $j$ corresponds to the $j\mathrm{th}$ excited state, $\rho^j_0 :=
\rho^j_\mathrm{SFDM}(0)$ is the central density and the radius $R$ of
the SF configuration is fixed through the condition
$\rho^j_\mathrm{SFDM}(R)=0$, which implies
\begin{equation}
	k_j = \frac{j \pi}{R} ; \,\,  j=1,2,3,...
\label{k_j}
\end{equation}
Following the usual interpretation, for $j$=1 the solution has no nodes
and it is usually associated to the ground state; for $j$=2, the SF
solution has one node and is interpreted as the first excited state;
larger $j$'s correspond to higher excited states.  Notice that this
solution presents a flat central profile.  In the Newtonian
approximation, the mass distribution of the state $j$ is given by
\begin{equation}
   M^j_\mathrm{SFDM}(r) = \frac{4 \pi \rho_0}{k_j^2} \frac{r}{2}
   \left[ 1- \frac{\sin(2 k_j r)}{(2 k_j r)} \right] .
\label{T-mass}
\end{equation}

  Since the field equation is linear, when the field rolls to the new
minimum of the potential, a more general case corresponds to a
superposition of the different modes of the SF, i.e. different
combinations of excited states that are coupled through the
gravitational potential that they generate.  For the general case, the
total density $\rho_\mathrm{SFDM}$ can be written as the sum of the
densities in the different states \citep{Robles:2013}:
\begin{equation}
 \rho_\mathrm{SFDM}(r) = \sum_j \rho^j_0 \left[ \frac{\sin(j \pi r
 / R)}{(j \pi r / R)} \right]^2 .
\label{densitytotal}
\end{equation}
Notice that in our case the radius $R$ is the same for all the excited
states and we define it as the radius of the SFDM halo.

  There have been previous works that explore the particular features of
the oscillations in the density profile trying to distinguish between
SFDM and CDM from the observations, including fitting rotation curves of
galaxies \citep{urena-bernal10,Robles:2013,Martinez-Medina:2014hca,
Bernal:2017}, the size of the Einstein radius of lensed galaxies by
strong gravitational lensing \citep{Robles-lensing:2013}, the production
of tidal substructures like shells and rings around massive galaxies
\citep{Robles-Medina:2015}, etc. We extend the study to show that in
galaxy clusters, we also have small oscillations due to the
superposition of all the states. However, the oscillations are so small
that they fall within the data uncertainties. Better data is required to
distinguish between the models.

\section{Clusters of galaxies: Data analysis}
\label{cluster-profiles}

\subsection{Navarro-Frenk-White profile}
\label{nfw-section}

  The discovery of the need of dark matter in clusters of galaxies comes
from the 1930s, when it was postulated in order to explain the
observations of the kinematics of the galaxies, in particular the Coma
\citep{Zwicky:1937} and the Virgo Clusters \citep{Smith:1936}. The
dynamics in these systems is dominated by the DM, which is about 90\% of
the total matter content, meanwhile the baryonic component is only 10\%
(from which the dominant component is the gas, about 9\%, in the hot
intracluster medium, ICM).  Assuming the clusters are dark matter dominated,
CDM $N$-body simulations provide the `universal' NFW profile that we use
to compare with the SFDM profiles.  We fit the same data of
\citet{Vikhlinin:2006} with the NFW profile given by \citep{nfw-1997}
\begin{equation}
	\rho_\mathrm{NFW}(r) = \frac{\rho_s}{(r/r_s)(1+r/r_s)^2} ,
\label{rho-nfw}
\end{equation}
whose mass profile is
\begin{equation}
  M_\mathrm{NFW}(r) = 4 \pi \rho_s r_s^3 \left[ - \frac{r}{r+r_s} + \ln
  \left( 1 + \frac{r}{r_s} \right) \right],
\label{nfw-mass}
\end{equation}
where $\rho_s$ is related to the density of the Universe at the moment
the halo collapsed and $r_s$ is a scale radius. The shape of the radial
density profile is characterised by a change in slope $\alpha =
\mathrm{d}\log \rho / \mathrm{d} \log r$, from $\alpha \approx -1$ in
the inner regions to $\alpha \approx -3$ at large radii
\citep{nfw-1997}. The profile is characterized by the concentration
parameter $c_\Delta:=r_\Delta/r_s$, evaluated at $r_\Delta$, the radius
where the density is $\Delta$ times the critical density of the
Universe. These concentrations are strongly correlated to the halo
formation epoch, which can be obtained from simulations. The values of
the concentration from the fits can then be compared with the expected
theoretical CDM values. 

  In clusters however, higher resolution simulations suggest that the
inner slope asymptotes between $\rho \sim r^{-1} - r^{-1.5}$, equal or
steeper than the original NFW profile \citep{Fukushige:1997, Moore:1999,
Jing:2000, Ghigna:2000}.  This apparent discrepancy between the CDM
simulations and the clusters' observations is generally attributed to
the physics of the baryonic matter, which contracts the DM increasing
the inner slope of the density profile.  It is expected that at the
innermost regions of galaxy clusters, the baryonic matter becomes a
significant component that we have to take into account in order to
understand the dynamics on these systems.  Knowing that baryons tend to
cool and collapse, the presence of baryons at the centres contributes to
steepen the density profile. Nevertheless, we are interested in the DM
distribution in the whole cluster so in this study we decided to fit
only data outside the brightest galaxy; a more in-depth analysis of the
central component will be given in a future work.

\subsection{Mass profile from X-ray observations}
\label{fit-observations}

  The mass distribution in clusters of galaxies can be determined by many
methods. The X-ray observations of the hot ICM is used to derive the
dynamical masses assuming the clusters are gravitationally bound
structures close to hydrostatic equilibrium, thus the gravitational
potential does not change considerably in a sound crossing time
\citep[see e.g.][]{Sarazin:1988}.  There are two observables from the
intracluster gas, the projected temperature and the X-ray surface
brightness, that can be modeled with three-dimensional gas density
$\rho_\mathrm{g}(r)$ and temperature $T(r)$ profiles.

  Under the hydrostatic equilibrium hypothesis, the spherically symmetric
hydrodynamic relation from the collisionless isotropic Boltzmann equation
in the weak field limit is \cite[see e.g.][]{binney-tremaine}
\begin{equation}
   \frac{\mathrm{d}\left[ \sigma^2_r \rho_\mathrm{g}(r) \right]}
   {\mathrm{d}r} + \frac{\rho_\mathrm{g}(r)}{r} \left[ 2 \sigma_r^2 -
   \left(\sigma^2_{\theta} + \sigma_{\varphi}^2 \right) \right] =
   - \rho_\mathrm{g}(r) \frac{\mathrm{d} \Phi(r)}{\mathrm{d}r} ,
\label{boltzmann}
\end{equation}
where $\Phi$ is the gravitational potential and $\sigma_r$,
$\sigma_{\theta}$ and $\sigma_{\varphi}$ are the gas velocity dispersions
in the radial and tangential directions, respectively. For relaxed
clusters in general, the gas velocities are isotropically distributed
\citep[e.g.][]{Sarazin:1988}, thus we have $\sigma_r = \sigma_{\theta} =
\sigma_{\varphi}$.

  For an ideal gas, the radial velocity dispersion $\sigma_r(r)$ is
related to the pressure profile $P(r)$, the gas mass density
$\rho_{\mathrm{g}}(r)$ and the temperature profile $T(r)$ by
\begin{equation}
   \sigma_r^2(r) = \frac{P(r)}{\rho_{\mathrm{g}}(r)} =
	\frac{k_{\mathrm{B}} T(r)}{\mu m_p},
\label{pressure-rhog}
\end{equation}
where $k_{\mathrm{B}}$ is the Boltzmann constant, $\mu=0.5954$ is
the mean molecular mass per particle for primordial He abundance and
$m_\mathrm{p} = 938 \,\mathrm{MeV}/c^2$ is the mass of the proton. Direct
substitution of~\eqref{pressure-rhog} into~\eqref{boltzmann}
yields the gravitational equilibrium equation
\begin{equation}
   - \frac{\mathrm{d} \Phi(r)}{\mathrm{d}r}
   = \frac{k_{\mathrm{B}} T(r)}{\mu m_p r}
   \left[ \frac{\mathrm{d} \ln \rho_\mathrm{g}(r)}{\mathrm{d}\ln r} +
   \frac{\mathrm{d} \ln T(r)}{\mathrm{d}\ln r} \right] .
\label{Beq-T-rho}
\end{equation}
From the last equation, the dynamical mass $M_\text{dyn}(r)$ of the
system is determined by
\begin{equation}
   M_{\mathrm{dyn}}(r) := - \frac{r^2 a_c}{G}
		    = - \frac{k_{\mathrm B} T(r)}{\mu m_p G} r \left[ \frac{
			\mathrm{d} \ln \rho_\mathrm{g}(r)}{\mathrm{d}\ln r} +
			\frac{\mathrm{d} \ln T(r)}{\mathrm{d}\ln r}
			\right] ,
\label{Newt-mass}
\end{equation}
where the right-hand side can be obtained using observational data. The
total cluster mass distribution is
\begin{equation}
   M_\mathrm{dyn}(r) = M_\mathrm{tot}(r) := M_\mathrm{bar}(r) +
   M_\mathrm{DM}(r) ,
\label{M-tot}
\end{equation}
where $M_\mathrm{bar}$ is the baryonic mass of the system and
$M_\mathrm{DM}$ the dark matter mass.  Thus, through the observed gas
density and temperature profiles, we can determine the total mass 
distribution of the cluster.

\subsection{Clusters' sample and data reduction}
\label{chandra-sample}

  For this work, we use the 13 clusters of galaxies from the
\textit{Chandra} X-ray Observatory analyzed by \citet{Vikhlinin:2005,
Vikhlinin:2006}, as a representative sample of low-redshift
($z\sim0.01-0.2$), relaxed clusters, spanning a range in temperatures of
$0.7-9$ keV (an energy band optimal for the signal-to-noise ratio in
\textit{Chandra}) and total masses $M_{500}$\footnote{The mass at the
radius $r_{500}$, where the density is $\rho(r_{500})=500
\rho_\mathrm{crit}$, with $\rho_\mathrm{crit}$ the critical density of
the Universe.}$\sim (0.5-10) \times 10^{14} M_\odot$, with very regular
X-ray morphology and weak signs of dynamical activity. In some cases,
data from the \textit{ROSAT} satellite was used to independently model the gas
density, strengthening the derived measurements. The observations are of
sufficient quality in \textit{Chandra} exposure times and extend up to
$0.5$ $r_{200}$ for all the clusters, which is a large fraction of the
virial radius $r_{200}$, thus they can obtain reliable measurements of
the cluster properties. As explained below, the gas density and
temperature profiles include several free parameters that are obtained
through Markov Chain Monte Carlo (MCMC) simulations until the optimal fit
is reached; the reliability of the modeling approach has been tested by
applying this procedure to high-resolution simulations of galaxy clusters
where the gas and temperature profiles can be independently known
directly from the simulation outputs \citep[see][] {Nagai:2007}.

  The hot gas haloes of massive galaxy clusters are intense X-ray
emitters. The observed keV temperatures are consistent with a fully
ionized ICM that mainly emits Bremsstrahlung radiation, but some line
emission by the ionized heavy elements might be found. The radiation
generated by these mechanisms is proportional to the emission measure
profile, $n_\mathrm{p} n_\mathrm{e}(r)$.  \citet{Vikhlinin:2006} introduced the following
modification to the standard $\beta$-model \citep{cavaliere1978} to
describe the emission measure and reproduce the observed features for
the 13 X-ray clusters:
\begin{equation}
\begin{split}
   n_p n_e(r) =& \frac{n_0^2 (r/r_c)^{-\alpha}}{(1+r^2/r_c^2)^{3 \beta -
   \alpha/2}} \frac{1}{(1+r^\gamma/r_{s_2}^\gamma)^{\epsilon/\gamma}} \\
	&+ \frac{n_{02}^2}{(1+r^2/r_{c_2}^2)^{3 \beta_2}} ,
\end{split}
\end{equation}
which includes two $\beta$-components in order to fit the profiles from
inner to outer radii.  All the clusters can be fitted adequately with a
fixed $\gamma = 3$ and the restriction $\epsilon < 5$ to exclude
non-physical sharp density breaks. The nine parameters ($n_0$, $r_c$,
$r_{s_2}$, $\alpha$, $\beta$, $\epsilon$, $n_{02}$, $r_{c_2}$ and $\beta_2$)
for the best-fitting emission measure profile for the clusters can be found
in table 2 of~\citet{Vikhlinin:2006}.

  From the emission measure profile, the gas density can be found taking
into account the primordial He abundance and the relative metallicity
$Z = 0.2 Z_\odot$, so
\begin{equation}
   \rho_\mathrm{g}(r) = 1.624 m_\mathrm{p} \sqrt{n_\mathrm{p}
   n_\mathrm{e} (r)} .
\label{rho_gas}
\end{equation}

  To have an estimation of the total stellar component, we use the
empirical relation between the stellar and the $M_{500}$ mass by
\citet{lin-vikhlinin12}:
\begin{equation}
   \frac{M_{\mathrm{stars}}}{10^{12} M_\odot} = (1.8 \pm 0.1) \left(
	\frac{M_{500}} {10^{14} M_\odot} \right)^{0.71 \pm 0.04} .
\label{stellar-mass}
\end{equation}
However, the stellar contribution to the total mass is $\sim 1\%$, so we
simply estimate the baryonic mass with the mass of gas and in the
following we neglect the stellar mass.

  The gas observations in these systems are generally well approximated
by an isothermal density profile decreasing as $\rho \sim r^{-2}$, which
is expected for a system in equilibrium. However, all the projected
temperature profiles show a broad peak near the centres, with a
temperature decline toward the cluster centres, followed by an
approximate constant temperature region and decreasing temperatures at
large radii. To model the observed features, \citet{Vikhlinin:2006} used
the following three-dimensional temperature profile:
\begin{equation}
  T(r) = T_0 \frac{x+T_\mathrm{min}/T_0}{x+1}
  \frac{(r/r_t)^{-a}}{\left[ 1 + (r/r_t)^b \right]^{c/b}} ,
\label{T-prof}
\end{equation}
where $x := \left( r/r_\mathrm{cool} \right)^{a_\mathrm{cool}}$.  We
refer the reader to table 3 of \citet{Vikhlinin:2006} for the resulting
8 best-fitting parameters ($T_0$, $r_t$, $a$, $b$, $c$, $T_\mathrm{min}$,
$r_\mathrm{cool}$ and $a_\mathrm{cool}$) for this temperature profile for
the 13 clusters of galaxies. To fit the temperature profiles, they
excluded from the fit the central regions ($r<r_\mathrm{min}$; see
Table~\ref{table1}) in order to avoid the multiphase gas at small radii,
substructures displaying active galactic nucleus (AGN) and the innermost
regions where the temperature profile does not describe well the observed
temperatures. In general, the central regions in clusters are not
spherically symmetric or in hydrostatic equilibrium, which could result
in a wrong estimation of the mass within these regions, where the most
massive brightest cluster galaxy (BCG) is dominant (up to $10^{12} M_\odot$ within
$r_\mathrm{min}$). We fit the dark matter models taking into account the
dark matter and gas only outside the cut-off radius $r_\mathrm{min}$ for
each cluster, excluding the BCG galaxy from the analysis; this leaves us
with a smaller baryon-to-dark matter mass fraction, however, for larger
radii outside the BCG galaxy the DM becomes the dominant component and
the total stellar contribution becomes only $\sim1\%$. We leave a
detailed study of the central cluster regions for a future work.

  The total dynamical masses, obtained with the use of
equation~\eqref{Newt-mass} from the derived gas densities~\eqref{rho_gas} and
temperature profiles~\eqref{T-prof}, for the 13 \textit{Chandra} X-ray
clusters of galaxies, were kindly provided by A.~Vikhlinin, along with
the $68\%$ confidence levels (CL) from the Monte Carlo simulations used
to fit the parameters of the models. We used the provided data and the
$68\%$ CL as the values to fit our dark matter models. We describe the
fitting method in the next subsection.

\subsection{Statistical calibration method}
\label{mcmc-method}

  In order to constrain the model parameters, we consider the likelihood
function $\mathcal{L}({\bf p})$, where ${\bf p}$ is the set of
parameters and the likelihood function is given by
\begin{equation}
\mathcal{L}({\bf p}) = \frac{1}{(2 \pi)^{\mathcal{N}/2}
|{\bf C}|^{1/2}} \exp{\left ( - \frac{{\bf \Delta}^{T}
{\bf C}^{-1} {\bf \Delta}}{2} \right )} ,
\label{eq:likelihood}
\end{equation}
with $\mathcal{N}$ the number of data points for each cluster and
${\bf \Delta}$ a column vector defined as
\begin{equation}
  {\bf \Delta} = M_\mathrm{dyn}(r) - M_\mathrm{model}(r,{\bf p}) ,
\label{Delta:Cl}
\end{equation}
where $M_\mathrm{dyn}$ is the best-fitting value obtained in
\citet{Vikhlinin:2006} and $M_\mathrm{model}$ is the corresponding mass
for a given dark matter model (CDM, SFDM, BECDM), while ${\bf C}$ is a
diagonal matrix.  In order to sample the parameter space, we used the
MCMC method~\citep{Gamerman:1997} with two
chains and checked the convergence with the Gelman-Rubin convergence
criterion ($R-1<0.1$) \citep{Gelman-Rubin:1992}. The best-fitting values and
confidence ranges are then calculated for all the parameters from the
resulting histograms.

  For each cluster we used the best-fitting data from \citet{Vikhlinin:2006}
as our data points and the 68\% errors from their MCMC simulations as
`uncertainties' to run our own MCMC simulation.

\section{Results and discussion}
\label{discussion}

\subsection{SFDM fits}

  Our results for the 13 galaxy clusters obtained from the MCMC method
for the finite-temperature multistate SFDM profile~\eqref{densitytotal}
are summarized in Table~\ref{table1}. We used a halo configuration formed
by three excited states corresponding to the values that minimized the
$\chi^2_\mathrm{SFDM}$ errors. We found very good agreement with the
dynamical masses from \citet{Vikhlinin:2006}. 

\begin{table*}
\begin{tabular}{cccccccc}
\hline
Cluster & $r_\mathrm{min}$ & $r_\mathrm{out}$ & $M_\mathrm{gas}$$^{a}$ &
$M_\mathrm{SFDM}$$^{b}$ & $M_\mathrm{tot}$$^{b}$ & $M_\mathrm{dyn}$$^{a}$ &
$\chi^2_\mathrm{SFDM}$$^{b}$ \\
 & $(\mathrm{kpc})$ & $(\mathrm{kpc})$ & $(10^{14}\mathrm{M}_\odot)$ &
 $(10^{14} \mathrm{M}_\odot)$ & $(10^{14} \mathrm{M}_\odot)$ &
 $(10^{14} \mathrm{M}_\odot)$ & \\
\hline
USGC S152	& 20.3	& 297  & $0.01\pm0.001$ & $0.13\pm0.02$ & $0.14\pm0.02$  & $0.13\pm0.01$  & 0.95 \\
MKW 4 		& 72.2  & 648  & $0.07\pm0.001$ & $0.69\pm0.10$ & $0.76\pm0.10$  & $0.83\pm0.09$  & 0.40 \\
A262 		& 62.3  & 648  & $0.11\pm0.001$ & $0.70\pm0.05$ & $0.81\pm0.05$  & $0.87\pm0.06$  & 0.90 \\
RX J1159+5531 & 72.2& 681  & $0.08\pm0.002$ & $0.90\pm0.16$ & $0.98\pm0.16$  & $1.12\pm0.16$  & 0.23 \\
A1991 		& 40.2  & 751  & $0.16\pm0.001$ & $1.19\pm0.10$ & $1.35\pm0.10$  & $1.32\pm0.14$  & 1.02 \\
A383 		& 51.3  & 958  & $0.44\pm0.004$ & $3.17\pm0.45$ & $3.61\pm0.45$  & $3.17\pm0.27$  & 0.52 \\
A133 		& 92.1  & 1006 & $0.32\pm0.003$ & $2.92\pm0.30$ & $3.24\pm0.30$  & $3.36\pm0.27$  & 0.68 \\
A907 		& 62.3  & 1109 & $0.65\pm0.004$ & $4.42\pm0.32$ & $5.07\pm0.32$  & $4.87\pm0.29$  & 1.91 \\
A1795 		& 92.1  & 1223 & $0.70\pm0.005$ & $5.33\pm0.21$ & $6.03\pm0.22$  & $6.16\pm0.42$  & 6.25 \\
A1413 		& 40.2  & 1348 & $0.96\pm0.008$ & $6.99\pm0.43$ & $7.95\pm0.44$  & $8.16\pm0.60$  & 5.74 \\
A478 		& 62.3  & 1348 & $1.05\pm0.007$ & $8.20\pm0.89$ & $9.25\pm0.90$  & $8.18\pm0.81$  & 0.70 \\
A2029 		& 31.5  & 1348 & $1.10\pm0.015$ & $8.13\pm0.33$ & $9.23\pm0.35$  & $8.38\pm0.62$  & 6.22 \\
A2390 		& 92.1  & 1415 & $1.66\pm0.017$ & $8.79\pm1.05$ &$10.45\pm1.07$  &$11.21\pm0.76$  & 0.49 \\
\hline
\end{tabular}
\caption{Best-fitting mass results from the MCMC method for the SFDM model
(equation~\ref{M-tot}). The best-fitting parameters were obtained with the combination of
three states of the finite-temperature multistate SFDM model, as
detailed in Table~\ref{table2}. From left to right, the columns
represent: (1)~the name of the cluster, (2)~the minimum $r_\mathrm{min}$
(excluding the region of the BCG) and (3)~maximum
$r_\mathrm{out}$ radii for the mass integration, (4)~the total gas mass
$M_\mathrm{gas}(r_\mathrm{out})$, (5)~the derived total SF
mass $M_\mathrm{SFDM}(r_\mathrm{out})$ from the MCMC method, (6)~the
resulting total mass $M_\mathrm{tot}(r_\mathrm{out})$ for the SFDM
model, (7)~the total dynamical mass $M_\mathrm{dyn}(r_\mathrm{out})$
and (8)~the minimum $\chi^2_\mathrm{SFDM}$
error from the MCMC method. All errors shown are at $\pm 1\sigma$
confidence level from the MCMC method used.\newline
$^{a}$From \citet{Vikhlinin:2006}.\newline
$^{b}$The present work.}
\label{table1}
\end{table*}

  Table~\ref{table2} shows the fitting parameters of our multistate SFDM
configurations.  Notice that different clusters have different
combinations of excited states, (e.g. $1+3+5$, $1+4+6$, etc.),
consistent with the merger formation scenario, reflecting how the
clusters had different merging histories.  We found that the reported
combinations of states are the ones that best reproduce the mass data.
It is always possible to include more states but we found that the
improvement to the fit is not substantial and it can add unnecessary
extra degrees of freedom to the fit. Given that our main intention is to
show whether the SFDM profile agrees with observations at cluster
scales, we have decided to keep the minimum number of parameters that
fit the data within the 68\% uncertainties; interestingly, no more than
three states are needed. 

  It is important to note that in some cases, the configurations show
combinations of excited states only (e.g. $2+3+6$).  In these clusters
it is also possible to include the ground state (or other states), but
we found that its contribution is negligible with respect to the other
three dominant states. For example, for the cluster A1991, the
contribution of the ground state ($j=1$) to the total mass profile is
only $\sim 1\%$.  Such an extremely small contribution has a negligible
impact on the total shape of the mass profile; therefore, we did not
include it in the fit.  This fact does not imply the lack of a ground
state within the cluster or inside each individual galaxy in the
cluster, where its relative contribution could be larger; we stress that
it simply means that the overall DM distribution within the cluster is
dominated by the other excited states. 

\begin{figure*}
\centering
  \includegraphics[width=8cm]{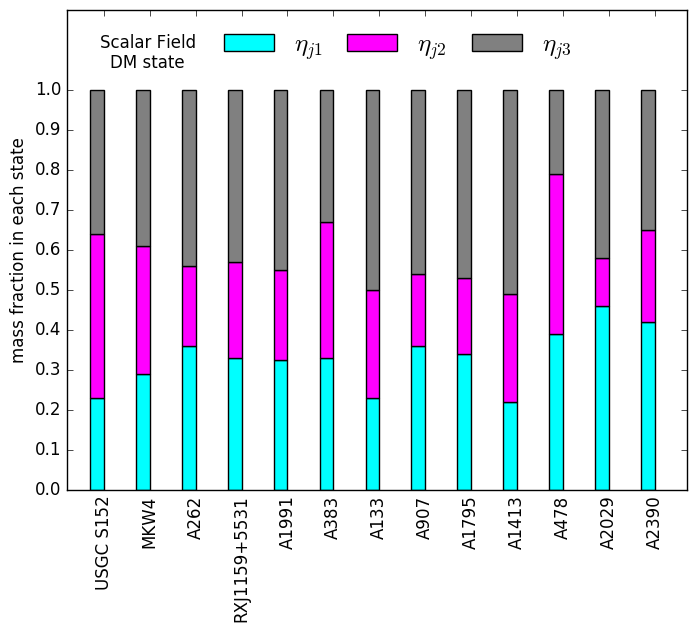}
  \includegraphics[width=8cm]{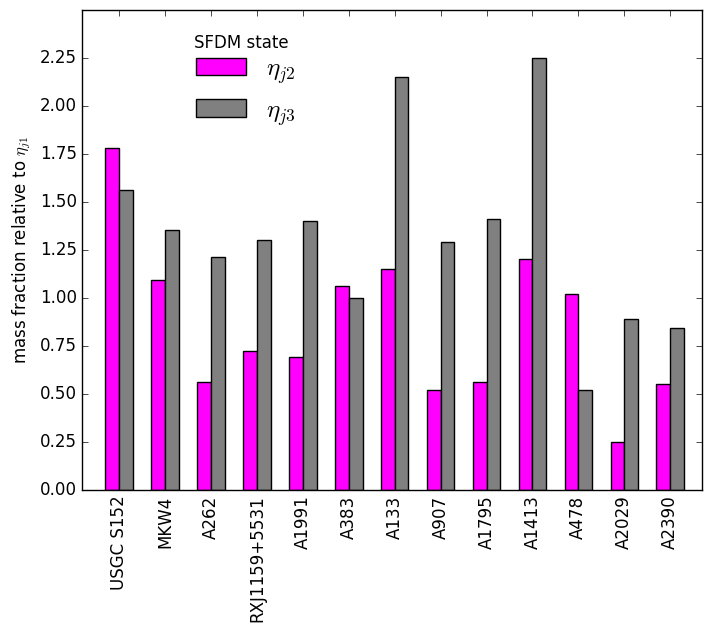}
  \caption{Left panel: stacked mass ratios of each state of the
multistate SFDM fit normalized to the total cluster mass from
equation~\eqref{densitytotal}.  In increasing-energy order, the minimum
energy state $\eta_{j_1}$ is plotted in cyan, the next dominant state
$\eta_{j_2}$ is in magenta and $\eta_{j_3}$ is shown in gray (colour
online). For different clusters, each state contributes differently to
the total mass reflecting the diverse formation history of each system.
Notice also that not all the clusters are composed of the same excited
states (see Table~\ref{table2}). Right panel: ratios of the mass in each
of the higher energy states with respect to the mass in the minimum
state ($\eta_{j_1}$). The mass ratios of most haloes are of order 1.3,
suggesting these haloes would be stable assuming the same stability
threshold of \citet{urena-bernal10}.}
\label{mass_ratios}
\end{figure*}

  In fact, we can compute the mass-ratios $\eta_{j_{1,2,3}} :=
M_{j_{1,2,3}} / M_\mathrm{SFDM}$, between the resulting total mass of
each dominant state, $M_{j_{1,2,3}}$, and the total SFDM mass of the
configuration, $M_\mathrm{SFDM}:=M_{j_1}+M_{j_2}+M_{j_3}$.  We report
these values in Table~\ref{table2}, and in Fig.~\ref{mass_ratios} we
see how much each state contributes to the total mass within
$r_\mathrm{out}$.  

  Notice that each state contributes in similar proportion to the total
mass, varying from 20-40\% for each state; also not all the clusters
have the same excited states in the fits, which is consistent with the
merger scenario of cluster formation. In fact, the variation can be
explained as follows: during the various mergers of individual haloes
that form the cluster, the different states of each halo mix and produce
the profile that we observe today.  It is possible that during the
merger process the SFDM particles were heated due to the
self-interactions of the SF and increased their energy yielding a larger
fraction of the DM particles in higher energy states. At the same time, 
gravitational cooling acts to bring particles of high energy to lower
modes of excitation of the field. The net effect of these two processes
changes as the cluster evolves and more mergers take place, increasing
the population of the excited states and making the amplitude of the
oscillations more pronounced. In the end, some fraction of the excited
states settles in the outskirts of each galaxy haloes living within the
cluster, but the rest still remains inside the deeper gravitational
potential of the cluster, as we observe it today. 

  Moreover, from Fig.~\ref{mass_ratios}, we observe that the
intermediate state shows the largest variation for the different haloes. 
As described above, this could be the result of the bosons mixing in the
halo; during the mergers some higher energy ultra-light bosons could lose
energy and populate the next lower level. At the same time, the bosons in
the intermediate excited state can also fall to the minimum energy state
that has a larger population of bosons, depleting  $\eta_{j_2}$. Thus,
the intermediate state acts as the intermediary between bosons of higher
and lower energy, being subject to the largest variation in boson
population until the total system gets closer to the equilibrium.

  In the right-hand panel of Fig.~\ref{mass_ratios} we show the contribution
from higher to lower energy states in the fit. These ratios remain of
order one, except for A133, A1413 and USGC S152, which display a larger
fraction of the most energetic state. If we extrapolate the stability
threshold of $\eta \sim 1.3$ \citep{urena-bernal10} for our multistate
configurations, these high fractions imply that the three haloes are not
in equilibrium. Nevertheless, it is uncertain which is the true stability
threshold for multistate SFDM haloes with high energy nodes, which makes
difficult to conclude about the stability of the clusters. 

  On the other hand, if these configurations were below their stability
threshold of collapse, the derived profiles would remain stable for a
long time. Thus, studying the correlations between the number of excited
states that fit the mass of the clusters at various redshifts and the
number of merger events in each of these systems could provide a more
quantitative comparison of cluster formation and evolution between the
SFDM paradigm and the CDM model. To accomplish the task, we need to
perform numerical simulations of SFDM that can track the evolution of a
cluster for a Hubble time and perform a statistical study, but we leave
this endeavour for a future work.

\begin{table*}
\begin{tabular}{ccccccccc}
\hline
Cluster & States & $\rho_{j_1}$ & $\rho_{j_2}$ & $\rho_{j_3}$ & $R$ & $\eta_{j_1}$ & $\eta_{j_2}$ &
$\eta_{j_3}$ \\
 & ($j_1+j_2+j_3$) & $(10^6 \mathrm{M}_\odot \mathrm{kpc}^{-3})$ &
 $(10^6 \mathrm{M}_\odot \mathrm{kpc}^{-3})$ &
 $(10^6 \mathrm{M}_\odot \mathrm{kpc}^{-3})$ & $(\mathrm{kpc})$ &
\\
\hline
USGC S152	& 2+3+4	& $0.60\pm0.07$	& $2.39\pm0.27$	& $3.80\pm0.28$	& $319 \pm8$  & 0.23 & 0.41	& 0.36 \\	
MKW 4 		& 1+3+5 & $0.13\pm0.01$	& $1.29\pm0.10$ & $4.43\pm0.27$ & $622 \pm15$ & 0.29 & 0.32 & 0.39 \\
A262 		& 1+3+5 & $0.18\pm0.01$	& $0.92\pm0.08$	& $5.52\pm0.20$ & $600 \pm8$  & 0.36 & 0.20 & 0.44 \\
RX J1159+5531 &1+4+6& $0.09\pm0.01$	& $1.26\pm0.19$	& $5.05\pm0.49$ & $806 \pm19$ & 0.33 & 0.24 & 0.43 \\
A1991 		& 2+3+6 & $0.58\pm0.05$	& $0.90\pm0.09$	& $7.40\pm0.19$ & $750 \pm8$  & 0.32 & 0.22 & 0.45 \\
A383 		& 2+3+6 & $0.51\pm0.07$	& $1.27\pm0.16$	& $5.08\pm0.29$ & $1112\pm20$ & 0.33 & 0.35 & 0.33 \\
A133 		& 1+3+5 & $0.06\pm0.01$	& $0.73\pm0.08$ & $3.67\pm0.18$ & $1250\pm12$ & 0.23 & 0.27 & 0.50 \\
A907 		& 2+3+4 & $0.48\pm0.04$	& $0.60\pm0.06$	& $2.79\pm0.13$ & $1301\pm6$  & 0.36 & 0.18 & 0.46 \\
A1795 		& 1+2+4 & $0.19\pm0.01$	& $0.41\pm0.03$	& $4.20\pm0.06$ & $1150\pm2$  & 0.34 & 0.19 & 0.48 \\
A1413 		& 1+2+5 & $0.10\pm0.01$	& $0.48\pm0.04$ & $5.64\pm0.09$ & $1350\pm2$  & 0.22 & 0.27 & 0.51 \\
A478 		& 2+3+7 & $0.58\pm0.07$	& $1.40\pm0.03$	& $4.13\pm0.16$ & $1539\pm39$ & 0.39 & 0.40 & 0.21 \\
A2029 		& 2+5+8 & $0.58\pm0.02$	& $0.95\pm0.13$	& $8.90\pm0.21$ & $1700\pm2$  & 0.46 & 0.12 & 0.42 \\
A2390 		& 1+2+6 & $0.18\pm0.01$	& $0.39\pm0.04$	& $5.49\pm0.27$ & $1481\pm27$ & 0.42 & 0.23 & 0.35 \\
\hline
\end{tabular}
\caption{Best-fitting parameters estimation from the MCMC method for the
multistate SFDM model (equation~\ref{densitytotal}). The columns represent: (1)~the name of the cluster, (2)~the
three states configuration that best fit the mass data, (3-5)~the
best-fitting central densities $\rho_{j_{1,2,3}}$ for the corresponding
$j_{1,2,3}$ states ($\pm 1\sigma$ errors), (6)~the best-fitting parameter
$R$ corresponding to the radius of the SFDM halo ($\pm 1\sigma$ errors)
and (7-9)~the mass-ratios $\eta_{j_{1,2,3}}$ between the resulting
total mass of every state, $M_{j_{1,2,3}}$, with respect to the total
SFDM mass, $M_\mathrm{SFDM}:=M_{j_1}+M_{j_2}+M_{j_3}$ [column (5) of
Table~\ref{table1}]. Note that the intermediate states not shown, e.g.
the ground state $j=1$ in some cases, not necessarily do not contribute
to the best-fitting estimation, but are negligible with respect to the
three dominant ones. It is possible to use more states to improve the
fit, but three states are good enough.}
\label{table2}
\end{table*}

\subsection{Comparison with other profiles}

  To compare the finite-temperature SFDM model~\eqref{densitytotal} with
the NFW density profile~\eqref{rho-nfw}, we fit the NFW parameters
$\rho_s$ and $r_s$ for the 13 galaxy clusters applying the MCMC method
explained in Section~\ref{mcmc-method}. The best-fitting results are
shown in Table~\ref{table3}, with the corresponding minimized
$\chi^2_\mathrm{NFW}$ errors. In general, the $\chi^2_\mathrm{SFDM}$
errors are smaller than the $\chi^2_\mathrm{NFW}$ ones, which
demonstrates that the multistate SFDM provides a great description of
galaxy clusters, and in some cases it is even better than the NFW
profile. We also compute the resulting $c_{500}:=r_{500}/r_s$
concentration parameters (for $r_{500}$ the radius where the density is
500 times the critical density of the Universe) and compare with those
from \citet{Vikhlinin:2006}.  We find that the values are within the
accepted range from the CDM simulations \citep{Dutton:2014}.

\begin{table*}
\begin{tabular}{cccccccc}
\hline
Cluster & $\rho_s$$^{b}$ & $r_s$$^{b}$ & $r_{500}$$^{a}$ & $c(r_{500})$$^{b}$ & $M_\mathrm{NFW}$$^{b}$ &
$\chi^2_\mathrm{NFW}$$^{b}$ \\
 & $(10^6 \mathrm{M}_\odot \mathrm{kpc}^{-3})$ & $(\mathrm{kpc})$ & 
 $(\mathrm{kpc})$ &  & $(10^{14} \mathrm{M}_\odot)$ & \\
\hline
USGC S152	&	$7.83\pm0.42$	&	$45.7\pm1.2$ &	.... & ....				&	$0.11\pm0.01$	&	1.21	\\
MKW 4 		&	$1.71\pm0.13$	&	$154\pm7$	&	$634 \pm28$ & $4.11\pm0.37$	&	$0.66\pm0.11$	&	0.92	\\
A262 		&	$1.64\pm0.10$	&	$159\pm6$	&	$650 \pm21$ & $4.09\pm0.29$	&	$0.68\pm0.10$	&	2.85	\\
RX J1159+5531&	$1.00\pm0.03$	&	$217\pm4$	&	$700 \pm57$ & $3.23\pm0.22$	&	$0.85\pm0.07$	&	1.41	\\
A1991 		&	$2.12\pm0.10$	&	$174\pm5$	&	$732 \pm33$ & $4.20\pm0.31$	&	$1.21\pm0.13$	&	0.39	\\
A383 		&	$1.71\pm0.10$	&	$271\pm9$	&	$944 \pm32$ & $3.48\pm0.23$	&	$3.13\pm0.43$	&	1.30	\\
A133 		&	$2.08\pm0.15$	&	$221\pm8$	&	$1007\pm41$ & $4.56\pm0.35$	&	$2.52\pm0.40$	&	1.10	\\
A907 		&	$1.58\pm0.09$	&	$309\pm10$	&	$1096\pm30$ & $3.55\pm0.21$	&	$4.33\pm0.56$	&	0.91	\\
A1795 		&	$1.05\pm0.03$	&	$411\pm7$	&	$1235\pm36$ & $3.00\pm0.14$	&	$5.78\pm0.39$	&	0.15	\\
A1413 		&	$1.19\pm0.05$	&	$408\pm11$	&	$1299\pm43$ & $3.19\pm0.19$	&	$7.02\pm0.74$	&	0.09	\\
A478 		&	$0.93\pm0.04$	&	$505\pm15$	&	$1337\pm58$ & $2.65\pm0.27$	&	$8.59\pm0.93$	&	5.01	\\
A2029 		&	$1.84\pm0.05$	&	$351\pm7$	&	$1362\pm43$ & $3.89\pm0.20$	&	$7.83\pm0.56$	&	0.23	\\
A2390 		&	$0.56\pm0.03$	&	$650\pm20$	&	$1416\pm48$ & $2.18\pm0.14$	&	$9.10\pm1.07$	&	0.28	\\
\hline
\end{tabular}
\caption{Best-fitting parameters estimation from the MCMC method for the NFW
density profile (equation~\ref{rho-nfw}). The columns represent: (1)~the name of the cluster, (2)~the
derived density parameter $\rho_s$, (3)~the derived characteristic
radius $r_s$, (4)~the observational $r_{500}$ radius, (5)~the resulting
concentration parameter $c(r_{500})$, (6)~the total DM mass from the NFW
profile, $M_\mathrm{NFW}(r_\mathrm{out})$ and (7)~the minimum
$\chi^2_\mathrm{NFW}$ error from the MCMC method. All errors shown are
at $\pm 1\sigma$ CLs from the MCMC method used.\newline
$^{a}$From \citet{Vikhlinin:2006}.\newline
$^{b}$This work.}
\label{table3}
\end{table*}

\begin{figure*}
\centering
  \includegraphics[width=17cm]{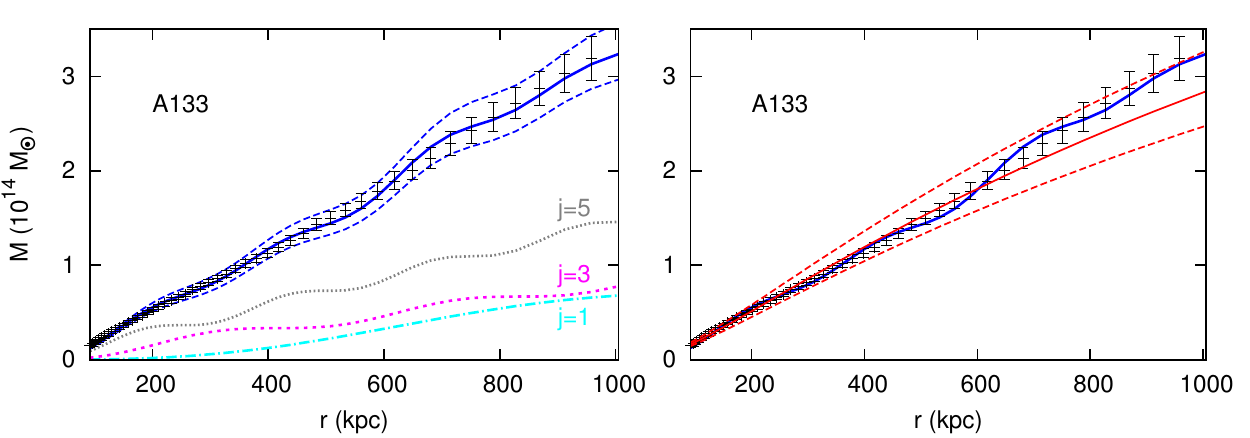}
  \caption{Left panel: dynamical mass vs. radius for the galaxy cluster
  A133, for the finite-temperature SFDM model. The points with uncertainty
  bars are the dynamical masses ($\pm 1\sigma$ errors) obtained by
  \citet{Vikhlinin:2006}. The blue solid line is the best fit obtained
  with the SFDM model and the blue dashed lines are the $1\sigma$ errors
  from the MCMC method. We show below the resulting masses for each of the
  three states that contribute to the total fit.  Right panel: the thin
  red line is the best fit from the NFW profile and the red dashed lines
  are the $1\sigma$ errors from the MCMC method. To compare, the thick
  blue solid line is the best fit from the multistate SFDM model (colour
  online).}
\label{plot_a133}
\end{figure*}

  The left-hand panel of Fig.~\ref{plot_a133} shows the fit corresponding to
the intermediate-mass cluster A133 for the SFDM model and its 68\%
confidence errors from the MCMC method, including the data with $1\sigma$
errors from \citet{Vikhlinin:2006}.  Included in the panel, we show the
resulting masses for the three states used to fit the dynamical mass for
each cluster. The right-hand panel of the same Figure shows our best fit for the
NFW profile and its 68\% error from the MCMC method, compared with the
best-fitting SFDM total mass for the same cluster A133.  We notice that for this
cluster the SFDM fits the dynamical mass better than the NFW profile, it
produces the characteristic oscillations at large radii and the total mass
falls within the data uncertainties at all times. Due to the combination of
excited states, the individual oscillations contribute differently to
the total profile and the wiggling behaviour is damped; the latter
effect is in addition to the intrinsic decay of the amplitude of the
oscillations, a consequence of the radial dependence of the density
profile (cf. equation~\ref{T-density}). Nevertheless, there is still a
residual that could be tested if data errors become smaller, although
the wiggling behaviour is expected to be more pronounced in isolated
galaxies where some cold gas could settle and better trace the
oscillations in the form of small density gradients
\citep{Robles-Medina:2015}, which could later seed star formation.

\begin{figure*}
\centering
  \includegraphics[width=17cm]{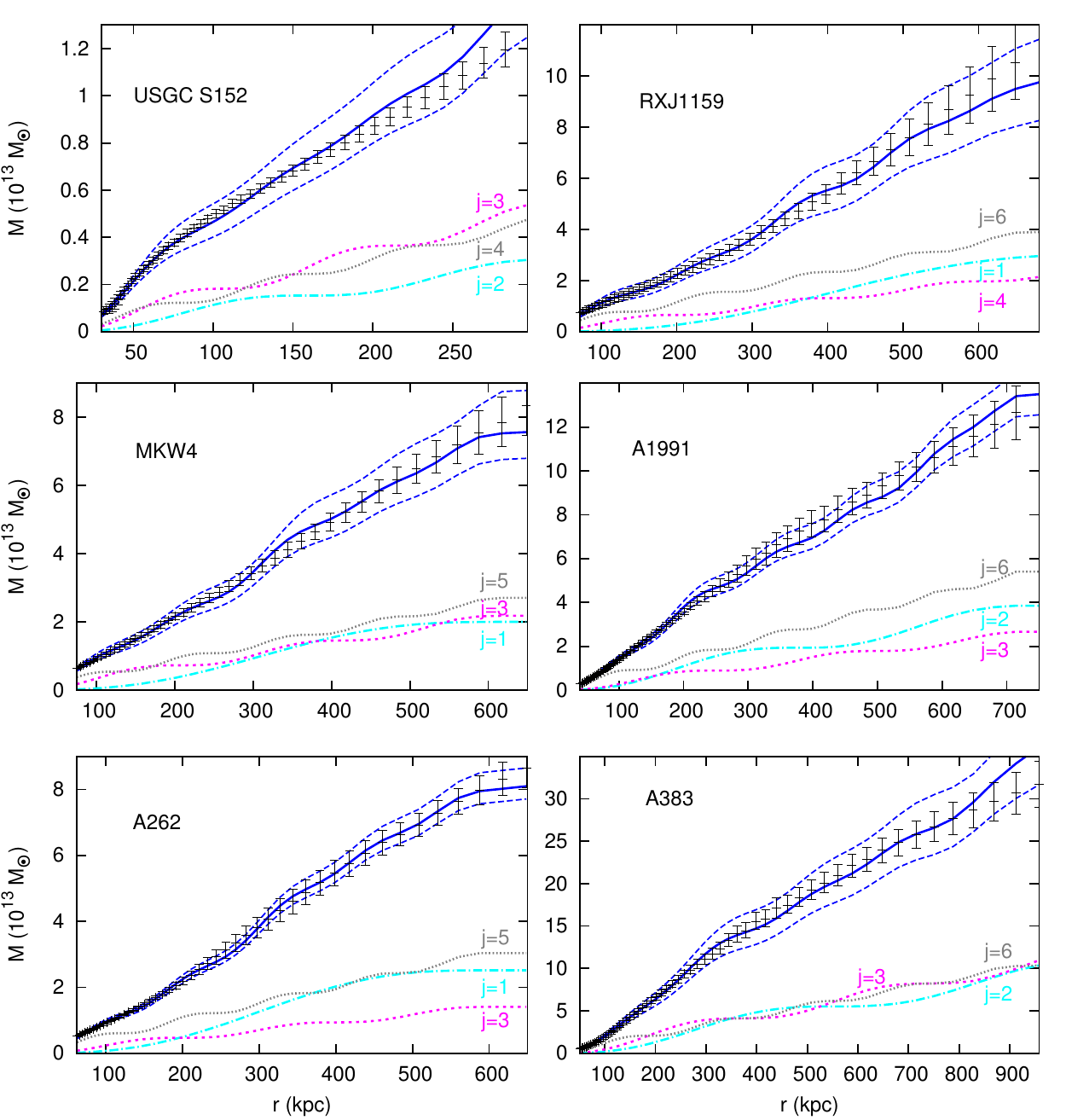}
  \caption{Same as Fig.~\ref{plot_a133}(left) for the first 6 clusters of
  galaxies of Table~\ref{table2}.}
\label{plot_data1-6}
\end{figure*}

\begin{figure*}
\centering
  \includegraphics[width=17cm]{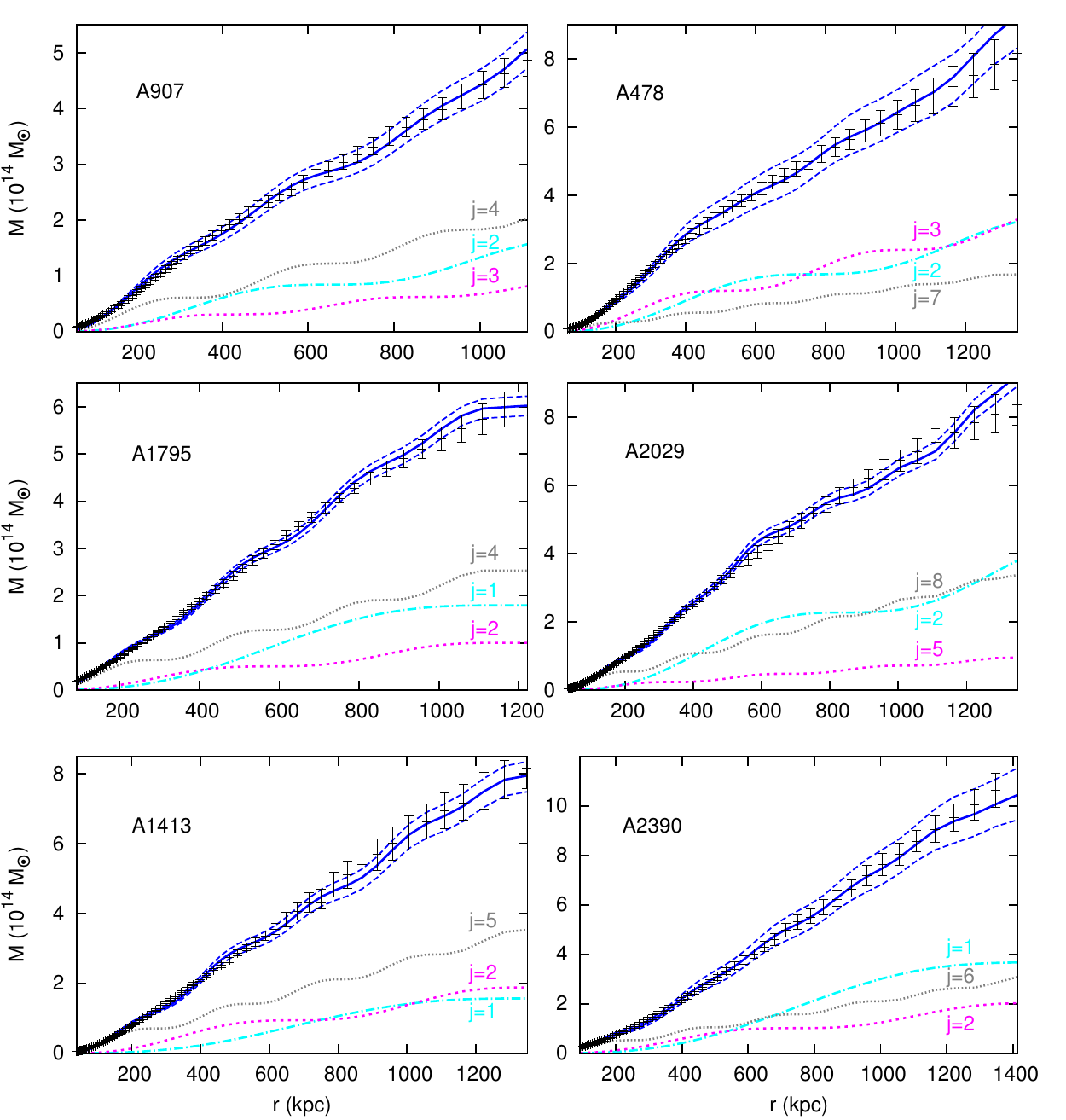}
  \caption{Same as Fig.~\ref{plot_a133}(left) for the last 6 clusters of
  galaxies of Table~\ref{table2}.}
\label{plot_data7-12}
\end{figure*}

\begin{figure*}
\centering
  \includegraphics[width=17cm]{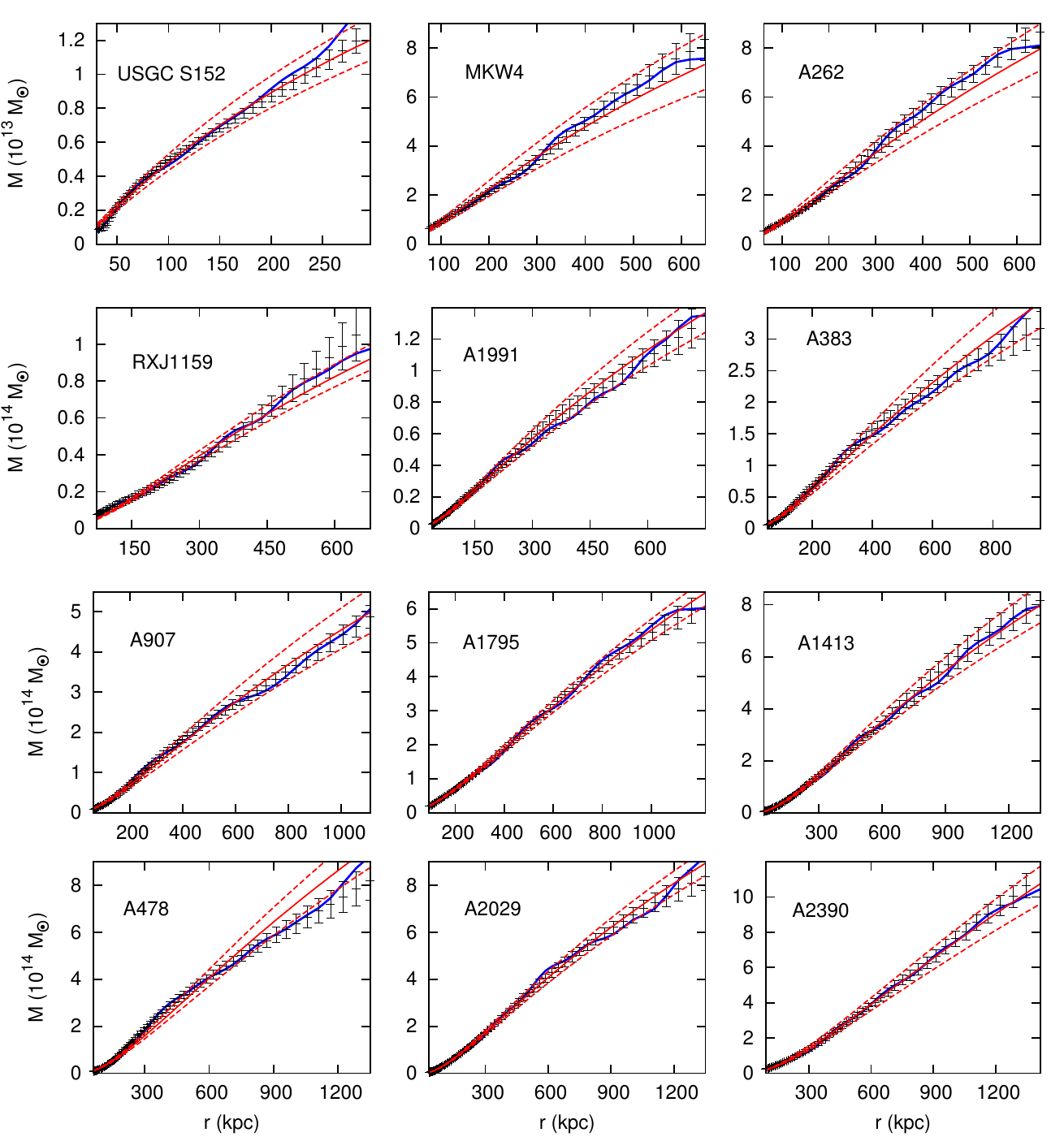}
  \caption{Same as Fig.~\ref{plot_a133}(right) for the last 12 clusters of
  galaxies of Table~\ref{table3}.}
\label{all-sfdm-nfw}
\end{figure*}

\begin{figure}
\centering
  \includegraphics[width=8cm]{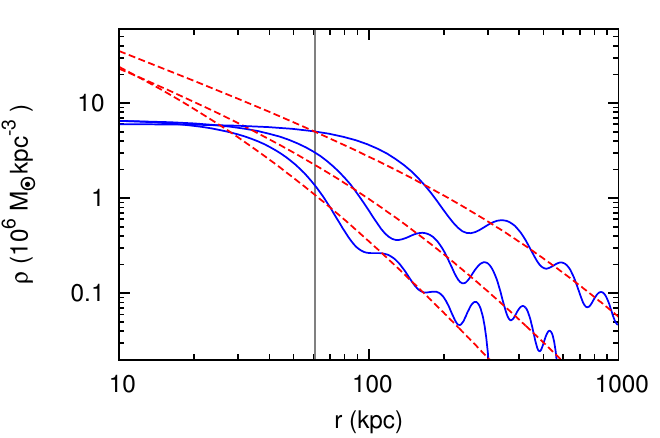}
  \includegraphics[width=8cm]{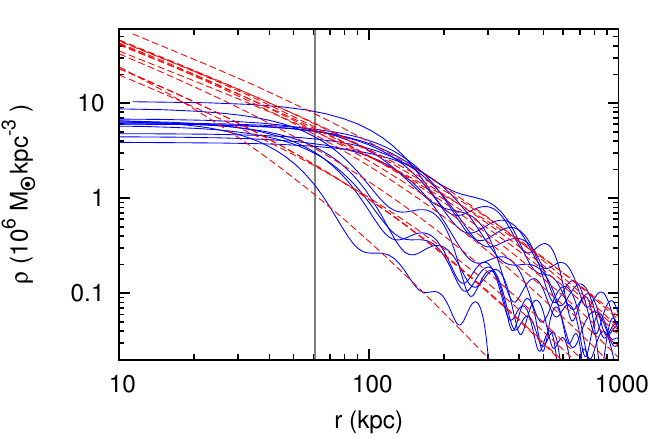}
  \caption{Top panel: density versus radius of three of our fits to the
  galaxy clusters. NFW profiles are shown in red dashed lines and the
  multistate SFDM fits are in blue solid lines (colour online). The
  oscillations of the SFDM model and the NFW-like dependence at
  intermediate radii are features also seen in SFDM simulations. The
  vertical line shows the average radius delimiting the excluded central
  regions in the fits; we extrapolate the density profiles within this
  radius to show the core nature of the SFDM profiles. Bottom panel:
  density profiles obtained for the 13 clusters. The same overlap is seen
  in all cases, different oscillations in distinct clusters are the result
  of the different excited states that appear on each fit.}
\label{densities-sfdm-nfw}
\end{figure}

  Comparing all the multistate SFDM fits (Figs.~\ref{plot_data1-6} and \ref{plot_data7-12})
with the NFW ones (Fig.~\ref{all-sfdm-nfw}), we observe that, in
general, the SFDM model follows the data at large radii equally well or
better than the NFW profile. For larger radii the oscillations become
smaller and in most cases the SFDM profile oscillates around the NFW one
for larger radii; the asymptotic behaviour for the SFDM solution is
however different [$\rho(r) \sim r^{-2}$ in SFDM as opposed to $\rho(r)
\sim r^{-3}$ in CDM], but for the range probed by the data they display a
similar decay in density. Remarkably, the overlap of the NFW and SFDM
profiles was also seen in recent simulations of mergers of SFDM haloes in
the ground state \citep{Schive:2014,Schive:2014hza, Schwabe:2016}. This
result motivated some authors to propose an ad hoc profile
composed of two pieces: a NFW profile for the outskirts of the haloes and
a core soliton-like density profile at the centres \citep{Schive:2014,
Marsh-Pop:2015}. In Fig.~\ref{densities-sfdm-nfw}, we show the density
profiles of the multistate SFDM compared to the NFW fits: in the upper
panel we plot the fits for three galaxy clusters to clearly demonstrate
that it is not necessary to assume such ad hoc approach: by
using the finite-temperature SFDM profile, which adds different
multistates, we can account for the NFW decay, limited to the radius
constrained by our current observations. The vertical lines in the figure
show the average radius delimiting the excluded central regions in the
fits, $\bar{r}_\mathrm{min}\sim 60$ kpc. In the bottom panel of the same
figure, we show that all our fits follow the same overlap with the NFW
profile, albeit the extra oscillations intrinsic to the wave nature of
the SF. Moreover, we extrapolate the fits to the central
regions of the clusters ($r<r_\mathrm{min}$), although in this region the
BCG can have an important contribution. We
notice that it is at this distance where the core-like nature of the SFDM
starts to be relevant \citep[see][for a discussion on the overlap between
the multistate SFDM model and the soliton+NFW profile obtained from the
high-resolution rotation curves of galaxies]{Bernal:2017}; adding the BGC
contribution would increase the central density in both models although
not necessarily in the same amount, which can put important constraints
on the models. This idea was applied by \citet{Elbert:2016} but to put
constraints to the self-interacting dark matter model; we leave a detail
analysis of the contraction in SFDM due to baryons for a future work.

\begin{figure}
\centering
  \includegraphics[width=8cm]{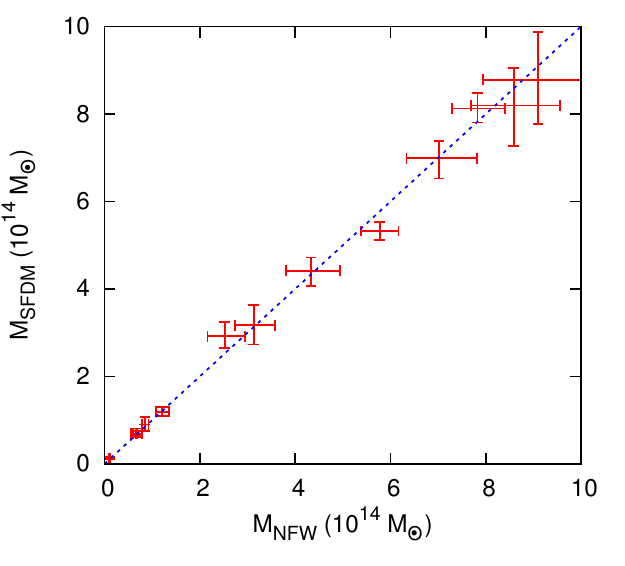}
  \caption{Total masses within $r_\mathrm{out}$ for the 13 clusters of
  galaxies obtained from the SFDM and NFW fits, at $1\sigma$ CLs. The blue
  dashed line (colour online) is a one-to-one linear relation. Both models yield similar
  total masses; however, their radial distributions are different as
  observed from Fig.~\ref{all-sfdm-nfw}.}
\label{plot-total-masses}
\end{figure}

  In Fig.~\ref{plot-total-masses}, we observe that the total DM mass
for the SFDM and NFW profiles is the same; we did not find any 
systematic trend in the SFDM profile regarding the total mass. This is
in fact reassuring as the total dynamical mass is the inferred value
required to explain the dynamics of the cluster, whereas the particular
radial distribution of DM can vary.  

\begin{figure*}
\centering
  \includegraphics[width=17cm]{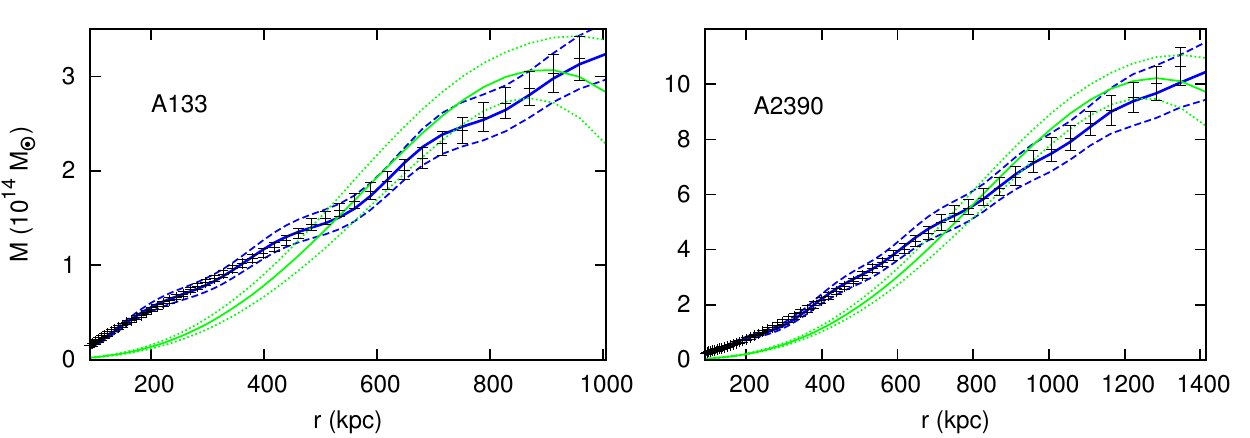}
  \caption{The figures show the best fit for the mass profiles of two of
  the clusters in our sample (green thin lines, colour online), A133 and A2390, and the
  $3\sigma$ errors from the MCMC method used (green dotted lines), for
  the zero-temperature BEC model in the TF approximation.  The best fit
  for the finite-temperature multistate SFDM profile (blue thick lines)
  and its $1\sigma$ errors (blue dashed lines) are shown for comparison
  reasons (colour online). For the first half of the data, the BEC
  profile underestimates the clusters' mass whereas for the other half
  the profile overestimates it. No oscillations are present in the BEC
  profile (solution of the GPP equation in the TF limit,
  equation~\ref{eq-1}). This is because it is only composed of bosons in the
  ground state, whose corresponding wavefunction has zero nodes.}
\label{plot-harko}
\end{figure*}

  In Fig.~\ref{plot-harko}, we compare the zero-temperature ground
state in the Thomas-Fermi limit (BEC profile) and the
finite-temperature multistate SFDM profiles, for two clusters in the
sample. It is noticeable how the BEC profile underestimates the mass of
the clusters for the first half of the data; in contrast, the SFDM
follows the data points all the way to the outermost radii.  The
discrepancy was also noted at the galactic scales with extended galaxies
\citep{Robles:2013}. From the fast decline of the BEC profile at large
radii, its associated rotation curve falls just after reaching the
maximum; this precludes the BEC profile to fit simultaneously the flat
part and the maximum of the rotation curves. In the case of dwarf
galaxies, where the data is constrained to the rising portion of the
rotation curve, the profile agrees with the data, but as observations
become available at larger distances, the BEC profile loses its
agreement. 

  In a recent work, \citet{Harko:2015} used an approximation based on
the statistical formulation to derive approximate thermal corrections to
the fully condensed BEC profile, caused by the collective excitations of
particles in the ground state. They showed that such effect would be
negligible today for BEC cluster haloes having most of the bosons in the
ground state. Our approach is quite different, we have used the SF 
potential from quantum field theory at finite temperature, which
can account for the excitations of the field, and apply the multistate
density profile obtained by \citet{Robles:2013} under the same formalism
to model the galaxy clusters DM distribution.  The finite-temperature
approach has been used in previous works and it has been shown to be
equivalent to the hydrodynamic approach \citep{suarez:2011}, which was
also shown to reproduce the large-scale observations \citep{suarez:2011,
Magana:2012,suarez:2015}.  In addition, \citet{Harko:2015} concluded
that the BEC model at zero temperature in the TF limit works better than 
their temperature-corrected profile to fit the total mass and radius of
a sample of 106 clusters of galaxies. Here we have found that the BEC
profile is in fact worse than the multistate profile. For this reason,
we believe our multistate profile can be used as a good approximation to
the full numerical solution, which was also suggested by
\citet{matos-urena07}.

\section{Summary and conclusions}
\label{summary}

  One of the alternative models to CDM is the SFDM model.  In this
model, the DM is a spin-0 scalar field
with a typical mass of $m \sim 10^{-22}\mathrm{eV}/c^2$ and a positive
self-interaction. This ultra-light boson is thought to form a
BEC which after the recombination
epoch behaves cosmologically as `cold' (pressureless and
non-relativistic) dark matter.  The equations that describe the scalar
field (SF) constitute the Einstein-Klein-Gordon system and, in the
non-relativistic limit, the BEC can be described by the Gross-Pitaevskii
and Poisson (GPP) equations.  The gravitationally bound solutions of the
GPP system of equations are interpreted as the dark matter haloes.

  Assuming spherical symmetry, the ground state (or BEC) solution
obtained in the Thomas-Fermi (TF) limit \citep{harko07}, where the SF
self-interactions dominate the SF potential ($V(\psi)\sim \psi^4$) in the
GPP system, can describe low-mass galaxies but is unable to keep the
flattening of the rotation curves for massive galaxies. However, it was
found that when finite-temperature corrections to the SF are taken into
account in the Klein-Gordon equation, there exist exact solutions of
excited states or, more generally, a combination of them, that provide an
accurate description of both small and large galaxies
\citep{Robles:2013}.

  In this work we explore the viability of these multistate SFDM
solutions at the galaxy clusters regime.  We fit the mass profiles of 13
\textit{Chandra} X-ray clusters of galaxies from \citet{Vikhlinin:2006}
using both the universal NFW profile predicted by CDM simulations and
the multistate density profile of the SFDM model. Additionally, we
compare it with the BEC solution in the TF limit. 

  We conclude that the analytic spherically symmetric SF configurations
obtained in the finite-temperature SFDM model \citep{Robles:2013} can
provide an accurate description of the DM mass distribution in galaxy
clusters on the range probed by the data.  As our main intention was to
test the overall consistency of the multistate SFDM profile and compare
it with the NFW and BEC profiles, we have left for a future work a more
detailed modelling of the region where the brightest cluster galaxy is
located. This region is dominated by the baryonic component and requires
a more in-depth analysis with numerical simulations addressing the
non-linear impact that baryons have on the core-like SFDM matter
distribution. Therefore, as a first approximation, we excluded the
central region of each cluster in the present article. 

  Our results suggest that the multistate SFDM profile agrees with the
data equally well as other empirical profiles currently used in the
literature \citep[see e.g.][for the soliton+NFW profile]{Schive:2014}, but
with the important difference that it is derivable from an underlying
theory and not just from an ad hoc profile.  In fact, the good
fits of the multistate haloes to the data at large radii suggest that
the approach of some authors to invoke an ad hoc profile to
parametrize the SFDM density profile obtained from numerical simulations
is not necessary.  The multistate SFDM profile can account for the
oscillating profile seen in simulations at large radii and also has an
overlap at intermediate radii with the NFW profile
(Fig.~\ref{densities-sfdm-nfw}). Also, it predicts the core-like
behaviour at the innermost radii in the clusters, in contrast to the
cuspy NFW profile. \citet{Bernal:2017} discussed the overlap between the
multistate SFDM model and the soliton+NFW profile obtained from the fits
of high-resolution rotation curves of galaxies.

  We found that galaxy clusters have different combinations of excited
states, reflecting their diverse formation history. The total profile 
follows the data at all radii and, in comparison to the NFW profile, in
some cases the SFDM is in better agreement especially at large radii
where the asymptotic decline of the profiles is different.

  From our comparison with the BEC profile at zero temperature
(Fig.~\ref{plot-harko}), we conclude that this profile is incapable of
fitting the entire mass distribution of the galaxy clusters at $3\sigma$
CL. We found similar results for other clusters.  Our
results complement the previously observed discrepancies at galactic
scales showing that the ground state in the TF limit is also
not a good description in the mass range of clusters, implying that a
purely self-interacting halo in the ground state is not adequate to model
very large systems. If the dark matter is indeed an ultra-light boson,
our results imply that the DM haloes of galaxies and galaxy clusters may
not be fully BEC systems. In contrast, agreement with observations at
different mass scales is achievable for multistate SFDM configurations.

  Finally, the oscillations in the SFDM profiles are within the data
uncertainties, making difficult their use to distinguish it from CDM.
Individual galaxies, particularly those in isolated environments where
tidal forces are smaller, are better candidates to look for these
wiggles. Being originated by small DM overdensities, the oscillations
could be seen either as low surface brightness gas overdensities in the
outskirts of the galaxies, or if the gas is cold and dense enough to
trigger star formation, they would create radial stellar gradients. Since
they damp with increasing radius, the largest effect would be given by
the first DM overdensity. These positive results from the multistate SFDM
fits motivate us to pursue a more detailed study of the centre of
clusters and obtain constraints on the model, especially by analysing
galaxy clusters where a core could be present \citep{Massey:2015,
Limousin:2016}.  Such study would require the addition of the baryonic
component to the total mass profile and likely in high-resolution
cosmological simulations in both CDM and SFDM models. Currently, the
studies of halo mergers in SFDM have been idealized due to intrinsic
code limitations \citep{Schive:2014,Mocz:2015,Schwabe:2016}.
Surprisingly, the analytic multistate solution indicates that it is an
interesting alternative profile that can be used as a fitting-function
for a large variety of gravitational systems and that the SFDM model
deserves further exploration.

\section*{Acknowledgements}

  We gratefully acknowledge Alexey Vikhlinin for providing the data of
the galaxy clusters used in this article.  This work was supported
by CONACyT M\'exico Projects (CB-2014-01 No. 240512, CB-2011 No. 166212
and 269652), Xiuhcoatl and Abacus clusters at Cinvestav and Instituto
Avanzado de Cosmolog\'{\i}a (IAC, http://www.iac.edu.mx/) collaboration
(I0101/131/07 C-234/07). We acknowledge CONACyT Fronteras 281.  TB and
VHR acknowledge economic support from CONACyT postdoctoral fellowships.

%%%%%%%%%%%%%%%%
% BIBLIOGRAPHY %
%%%%%%%%%%%%%%%% 
\bibliographystyle{mnras}
\bibliography{sfdm-clusters}

%%%%%%%%%%%%%%%%%%%%%%%%%%%%%%%%%%%%%
\bsp	% typesetting comment
\label{lastpage}
\end{document}